\documentclass[12pt]{article}
\usepackage{cite}
\usepackage[pdftex]{hyperref}
\parskip=\baselineskip
\title{}
\author{Davide Gaiotto}
\begin{document}

\begin{titlepage}
\begin{flushright}
hep-th/yymm.nnnn
\end{flushright}
\vskip 1.5in
\begin{center}
{\bf\Large{Monster symmetry and Extremal CFTs}} \vskip 0.5in {Davide
Gaiotto} \vskip 0.3in {\small{ \textit{School of Natural Sciences,\\
Institute for Advanced Study,\\ Einstein Dr., Princeton, NJ 08540}}}

\end{center}
\vskip 0.5in

\baselineskip 16pt
\date{}

\begin{abstract}
We test some recent conjectures about extremal selfdual CFTs, which
are the candidate holographic duals of pure gravity in $AdS_3$. We
prove that no $c=48$ extremal selfdual CFT or SCFT may possess
Monster symmetry. Furthermore, we disprove a recent argument against
the existence of extremal selfdual CFTs of large central charge.
\end{abstract}
\end{titlepage}
\vfill\eject

\tableofcontents
\section{Introduction}

The author of \cite{Witten:2007kt} argues that the holographic dual of pure (super)gravity in three dimensions should be a holomorphicaly factorized $c=24 k$ ($c=12k^*$) (super)conformal field theory  with as few low dimension (super)Virasoro primaries as possible.
Modular invariance alone prescribes a unique factorized torus partition functions with no Virasoro primary field of dimension smaller or equal to $\frac{c}{24}$. It is not known if an ``Extremal'' CFT with such a partition function actually exists for $k>1$, nor if it is unique: some non-trivial checks on
the existence of low-$k$ ECFTs are available
\cite{Gaiotto:2007xh,Witten:2007kt,Yin:2007gv}, but
\cite{Gaberdiel:2007ve} formulates a conjecture which implies the non-existence of $k>42$ ECFTs.
The only known examples, the $k=1$ ECFT and the $k^* = 1,2$ ESCFT, possess a very large discrete symmetry group: the most notable is the $k=1$ theory, which possesses two copies of the Monster discrete symmetry group, each acting on one chiral half of the theory. It is conjectured in \cite{Witten:2007kt} that the Monster group might be a symmetry of 3d quantum gravity in general. Also,
the tentative partition function for a $k^*=4$ theory also suggests
Monster symmetry \cite{Witten:2007kt}.

The purpose of this note is to test the above conjectures. In the
first section we prove that no $k=2$ ECFT may possess Monster
symmetry. Our strategy is simple: If a ECFT exists which has Monster symmetry, it must be
possible to define twist fields $\sigma_g$ and twisted partition
functions $Z_g =\mathrm{Tr}\,g\,q^{L_0}$ for each symmetry group element $g$
in the Monster. It turns out to be impossible to consistently build such
objects for group elements in the $2A$ conjugacy class of
the Monster group. The approach works as well for the $k^*=4$
theory, which is discussed in the second section. Our strategy does
not allow a clear cut conclusion about higher $k$ ECFTs. The partial
results we present here should still be useful in constraining the
symmetry structure of generic ECFTs. Additionally, in the last section we will present a counterexample to the conjecture in
\cite{Gaberdiel:2007ve}.

The motivation for this work is quite straightforward.
The only strategy available at this time to construct ECFTs with $k>1$ is direct conformal bootstrap. It is conceivable
that an ECFT may be defined as a $W$ algebra by a finite set of
OPEs for a choice of Virasoro primary fields. There are
some indications, for example, that a conformal bootstrap is possible for the $k=1$ Monster module. The dimension $2$, $3$, $4$, $5$ primaries all sit in single irreducible representations of Monster symmetry, and some normal ordered products of the dimension $2$ fields and their
derivatives do sit in the same irreps as the dimension $4$ and $5$
fields. If those normal ordered products are not null vectors, the singular part of the OPE of dimension $2$ and $3$ fields will involve only other dimension $2$ and $3$ fields and their normal ordered products. This defines a W-algebra. Because of Monster symmetry, the only unknown coefficients in the OPE are the overall normalization of the three point functions.
Associativity and unitarity of the W-algebra can be reduced to a set of Fierz identities for the appropriate Monster representations.
One may hope for a similar structure of OPEs in a higher $k$
ECFTs, which close on the fields of level from $k+1$ to $2k+1$.
The large number of fields involved even for very small $k$ makes it
clear that only a large discrete symmetry group will make the
conformal bootstrap possible in practice. This makes it important to
find some way to test for the existence of Monster symmetry in the
next simplest example, $k=2$, and possibly find a prescription of
which Monster irreps would contain the level $3$ to $5$ fields, to
jump-start a bootstrap procedure. Unfortunately our negative results
make this strategy unfeasible, unless a different candidate is found for a large symmetry group. On the other hand the counterexample presented in the last section of this note makes it clear that the existence of extremal CFTs is still a possibility for all $k$.

\section{$k=2$ ECFT}

A ECFT with a chiral symmetry group
$G$ has chiral $g$-twisted partition functions $Z_g(\tau) = \mathrm{Tr}\,g\,
q^{L_0}$ for each $g \in G$, which have interesting modular
properties. For example, the set of twisted partition functions
$t_g(\tau)$ for the the Monster module is quite famous, and is
identical to the set of genus zero hauptmodules. This fact was
instrumental in the discovery of the Monster module itself.
The modular group transforms the partition functions $Z_g(\tau)$ are
a set of more general $Z_{g^a,g^b}$, defined as a torus partition
function with $g^a$ and $g^b$ inserted at the two cycles of the
torus. The action of the modular group is quite intuitive:
\begin{equation}
S: (a,b)\to (b,-a) \qquad \qquad T: (a,b) \to (a,a+b)\end{equation}
If an anomaly is present for the chiral group symmetry, the modular group acts projectively, and extra phases appear in the transformation rules.
The coefficients of the twisted partition functions $Z_{g} =
Z_{e,g}$ must decompose into characters of the symmetry group
appropriately: if $Z(\tau) = \sum_R d_R f_R(\tau)$ then
$Z_{e,g}=\sum_R \chi_R(g) f_R(\tau)$. On the other hand the
coefficients of the partition functions of twisted sectors
$Z_{g^a, g^b}$ must decompose into characters of the stabilizer of
$g^a$ in the symmetry group.

The first constraint on the existence of a ECFT with Monster
symmetry is that such twisted partition functions must exist for all
the Monster group conjugacy classes, be compatible with each other
and satisfy certain positivity and integrality requirements. (The
untwisted partition function of a twisted sector $Z_{g,e}$ must have
positive integers as coefficients, up to a global ambiguous phase).
This is not quite enough to fix the partition functions uniquely. As
described in the appendix A, there is a standard recipe to build
candidates as ``twisted'' Hecke transforms of the hauptmodules
$t_g(\tau)$.
On the other hand the analysis of modularity will be enough to
determine the value of the dimension of the ground state in the $2A$
twisted sector of the $k=2$ ECFT. This value will turn out to be
inconsistent with the OPE of the twisted sector ground state with
itself.


The modular properties of a $Z_2$ twisted partition function are
described by the following diagram:
\begin{equation}
Z_g \Leftrightarrow_T Z_g \Leftrightarrow_S Z^g \Leftrightarrow_T
Z^g_g \Leftrightarrow_S Z^g_g
\end{equation}
The modular transformations above may in general be anomalous, so
that phases are generated through the transformations. An example of
this phenomenon is evident in the list of Hauptmodules $t_g$.
For example $t_{3C}$ is $Z_{3C} =
J(3\tau)^{\frac{1}{3}}$, the $S$ transform of it is
$Z^{3C}=J(\tau/3)^{\frac{1}{3}}$, which has a power expansion
involving powers $q^{\frac{n}{3}-\frac{1}{9}}$. Under $T^3$ this
partition function of the twisted sector goes back to itself as
expected, but with a phase of $\frac{2 \pi}{3}$.

The anomalous phases are constrained by the requirement that $S^2=1$
and $(ST)^3=1$. In certain cases, in particular for the case at hand
of a $Z_2$ symmetry, these two equations fix all phases up to some
discrete choices. Let us fix a few irrelevant phases by defining $Z^g$
as the $S$ transform of $Z_g$ and $Z_g^g$ as the $T$ transform of
$Z^g$. The $S$ transform of $Z_g^g$ should be $Z_g^g$ itself up to a
phase, and $S^2=1$ fixes the phase to be $\epsilon = \pm 1$. The $T$
transform of $Z_g^g$ should be a multiple of $Z^g$ again, up to a
phase $\alpha$. $(ST)^3=1$ acting over $Z_g$ gives $\epsilon
\alpha=1$.
Hence there are two possibilities: if $\epsilon=1$, $Z^g$ will have
a $q$ expansion involving integer and half integer powers of $q$; if
$\epsilon=-1$ $Z^g$ will have a $q$ expansion involving powers of
the form $q^{n \pm 1/4}$.
After the precise modular transformation rules are known, the set of
partition functions can be organized into a vector-valued holomorphic
modular form. Such forms are uniquely specified (up to a possible
constant term) by the set of all their polar terms, which are the
coefficients of the negative powers of $q$. For a unitary theory of
central charge $48$ and $\epsilon=1$ the polar terms are
\begin{equation}
Z_g = \frac{1}{q^2} + \frac{a_1}{q} + a_0 +\mathcal{O}(q) \qquad \qquad Z^g=\frac{b_{3/2}}{q^{3/2}} + \frac{b_1}{q} +
\frac{b_{1/2}}{q^{1/2}}+\mathcal{O}(1)\end{equation}
The constant term $a_0$ is a ``polar
term'' iff all the anomalous phases are zero.
If $\epsilon=-1$ the polar terms are
\begin{equation} Z_g=\frac{1}{q^2} +
\frac{a_1}{q}+\mathcal{O}(1) \qquad \qquad Z^g=\frac{b_{7/4}}{q^{7/4}} + \frac{b_{5/4}}{q^{5/4}} +
\frac{b_{3/4}}{q^{3/4}}+ \frac{b_{1/4}}{q^{1/4}}+\mathcal{O}(q^{3/4})\end{equation}.

A basis of modular forms for the $\epsilon=-1$ case can be readily
generated:
\begin{equation} \frac{E_6(2\tau)}{\eta(2 \tau)^{12}}=\frac{1}{q}+\mathcal{O}(1)\qquad \qquad S \left[\frac{E_6(2\tau)}{\eta(2 \tau)^{12}}\right]= -\frac{1}{q^{1/4}}+\mathcal{O}(q^{3/4})\end{equation}
\begin{equation} \frac{E_6(\tau)}{\eta(2\tau)^{12}}=\frac{1}{q}+\mathcal{O}(1)\qquad \qquad S\left[\frac{E_6(\tau)}{\eta(2\tau)^{12}}\right]=-\frac{64}{q^{1/4}}+\mathcal{O}(q^{3/4})\end{equation}
\begin{equation} 64\frac{E_6(\tau)\eta(2\tau)^{12}}{\eta(\tau)^{24}}=\mathcal{O}(1) \qquad \qquad S\left[64\frac{E_6(\tau)\eta(2\tau)^{12}}{\eta(\tau)^{24}}\right]=-\frac{1}{q^{3/4}} +
\frac{12}{q^{1/4}}+\mathcal{O}(q^{3/4})\end{equation}
Any other modular form with these modular
properties can be reproduced from its polar terms as a linear
combination of those three with coefficients polynomials of $J$.
Interestingly, for all $k$ we cared to test (up to $k=13$), and in
particular for $k=2$, if the polar terms of $Z_g$ are set to the
coefficients of the vacuum Virasoro module and the polar terms of
$Z^g$ are integer, the constant coefficient $a_0$ turns out to be a
multiple of $8$. On the other hand for $k<10$ the number of Virasoro
vacuum descendants of dimension $k$ is not a multiple of $8$. This
is a contradiction, and implies that no consistent anomalous $Z_2$
twisted partition function can be written for an ECFT of central
charge $24k$, $k<10$.

The non-anomalous case requires extra information from the OPEs.
Let us start with an instructive example: the OPE of $Z_2$ twist
fields in the case of the $k=1$ ECFT. The $2B$ twisted partition
function can be written in terms of the discriminant as
\begin{equation}
Z_{2B}=T_{2B}=\frac{\Delta(\tau)}{\Delta(2 \tau)}-24=\frac{1}{q}+276
q-2048 q^2+11202 q^3-49152 q^4+184024 q^5 + \mathcal{O}(q^6)
\end{equation}

The modular transform of it is the partition function of a $2B$
twisted sector.
\begin{equation}
Z^{2B}=2^{12} \frac{\Delta(\tau)}{\Delta(\tau/2)}-24=24+4096
\sqrt{q}+98304 q+1228800 q^{3/2}+10747904 q^2+\mathcal{O}(q^{5/2})
\end{equation}

Notice that the $24$ twisted sector ground states have dimension
one, and that their OPE is in the untwisted sector. The singular part must be

\begin{equation}
\sigma_i(z)\,\sigma_j(0) \sim \frac{\delta_{ij}}{z^2}
\end{equation}

Hence they are $24$ independent free bosons.
This is quite obviously consistent
with the fact that the $2B$ orbifold of the Monster CFT is the same
as the Leech Narain lattice model.

On the other hand the $2A$ twisted partition function is
\begin{equation}
Z_{2A}=T_{2A}=\frac{\Delta(\tau)}{\Delta(2
\tau)}+2^{12}\frac{\Delta(2\tau)}{\Delta(\tau)}-24=\frac{1}{q}+4372
q+96256 q^2+1240002 q^3 + \mathcal{O}(q^4)
\end{equation}

The modular transform of it is the partition function of a $2A$
twisted sector.
\begin{equation}
Z^{2A}=2^{12}\frac{\Delta(\tau)}{\Delta(
\tau/2)}+\frac{\Delta(\tau/2)}{\Delta(\tau)}-24=\frac{1}{\sqrt{q}}+4372
\sqrt{q}+96256 q+1240002 q^{3/2} + \mathcal{O}(q^2)
\end{equation}

The single $2A$ twist field must have OPE
\begin{equation}
\sigma(z)\,\sigma(0) \sim \frac{1}{z}
\end{equation}
and is a free fermion.

The first regular term in the OPE of a free fermion with itself is a
$c = \frac{1}{2}$ energy momentum tensor $T_{\sigma}$. It is a well
known fact that among the level two primaries of the $k=1$ theory
one can always pick a set of free-fermion energy momentum tensors.

It should be clear that this sort of consideration becomes very
constraining at higher $k$, where no level two primaries exist. For
example the $Z_2$ twist fields cannot have dimension $\frac{1}{2}$
or $1$, otherwise they would be free fermions and free bosons, and
generate extra energy momentum tensors with central charge
$\frac{1}{2}$ or one, while at level $2$ only a single energy
momentum tensor of $c=24k$ exists.
If the twisted sector ground state is a dimension $3/2$ field, the OPE with itself will contain the total energy momentum tensor, and will give rise to
a superconformal algebra.
As there are no dimension $1$ currents in the untwisted sector, this
algebra can have $N=1$ at most, hence there can only be up to one
twisted sector ground state of dimension $3/2$.
So we have learned that $b_{3/2} = b_1 =0$, $b_{1/2}
= 0$ or $1$.

The partition functions can be written in terms of polynomials of
$J$ acting on $1$, $\displaystyle{\frac{\eta(\tau)^{24}}{\eta(2
\tau)^{24}}}$ and $\displaystyle{\frac{\eta(2 \tau)^{24}}{\eta(\tau)^{24}}}$, and will be
\begin{equation}
Z_g = 1+\frac{1}{q^2}+(-4096+4096 b_{1/2}) q+(98580+98304 b_{1/2})
q^2
\end{equation}
And
\begin{eqnarray}
Z^g &=& \frac{b_{1/2}}{\sqrt{q}}+25-24 b_{1/2}+(196608+276 b_{1/2})
\sqrt{q}+ \nonumber\\ &+&(21495808-2048 b_{1/2}) q+(864288768+11202 b_{1/2}) q^{3/2}
\end{eqnarray}
Positivity of the coefficients of $Z^g$ also confirms $b_{1/2} = 0$
or $1$.
Correspondingly, the partition function $Z_g$ will be either
$T_{2B}^2 - 551$ of $J(2 \tau) +1$. The first case is a perfect
candidate for $Z_{2B}$. The second case is \emph{not} a good candidate for
$Z_{2A}$, as it gives zero as the character for the dimension $3$
primaries, but all the characters for $2A$ in the lower irreps of
the monster are positive numbers. Hence the $k=2$ theory cannot have
Monster symmetry

What can we say for higher values of $k$? Already at $k=3$ the
constraint is much weaker. Rough candidate partition functions can
be readily generated with a single spin field ground state of
dimension $3/2$, which forms a Super Virasoro algebra with the
energy momentum tensor. A specific candidate really shines through:
the one whose polar terms contain one spin field of dimension $3/2$
and one of dimension $2$. The coefficients decompose appropriately
in characters of the Monster, the coefficients of the modular
transform decompose neatly in dimensions of the centralizer of $2A$.
Before concluding that this partition function provides strong
evidence for Monster symmetry at $k=3$ though, the reader is invited
to check Appendix A, where a simple physically motivated
construction, the twisted Hecke transform, is described. The
construction makes it clear that there is a standard way to generate
candidate twisted partition functions for a theory with any possible
spectrum of polar untwisted states, such that all the partition functions
are automatically consistent with each other in the decomposition in
characters of the Monster, and such that their modular transforms
decompose nicely in dimensions and characters of the centralizer.
This fact is a direct consequence of the existence of the $k=1$
Monster module, so the existence of a nice candidate for a ECFT
twisted partition function is hardly a surprise. It is possible that
the requirements of positivity of the coefficients of partition
functions in twisted sectors will still have some strength, but we
will not pursue the matter further in this note.
Interestingly, our methods can be used to exclude such
deceptively beautiful candidates for $k>3$. Spin fields of dimension
$3/2$ or $2$ are not allowed anymore, as their normal ordered
product would be a non-existent non-Virasoro dimension $4$ untwisted
field. Twist fields of dimension up to $k/2$ can also be excluded,
as they would generate non-existent
W-algebras.\cite{Bouwknegt:1988sv}



\section{$k^*=4$ ESCFT}
A similar reasoning can be applied to the partition function of
extremal SCFTs.
The modular structure of twisted partition functions is slightly
more complicated. There are nine partition functions for a $Z_2$ symmetry $g$, with three
choices of spin structure and three of g-twist. There are two orbits
under the modular group.
The smaller orbit has size three, it involves the Ramond partition
function with a g insertion $Z^R_{\mathit{NS},g}$, the partition function in
the g-twisted NS sector with $(-1)^F$ inserted, $Z^{\mathit{NS},g}_R$, and
the g-twisted NS partition function with a g insertion
$Z^{\mathit{NS},g}_{\mathit{NS},g}$,

\begin{equation}
Z^R_{\mathit{NS},g} \Leftrightarrow_T Z^R_{\mathit{NS},g} \Leftrightarrow_S Z^{\mathit{NS},g}_R
\Leftrightarrow_T Z^{\mathit{NS},g}_{\mathit{NS},g} \Leftrightarrow_S Z^{\mathit{NS},g}_{\mathit{NS},g}
\end{equation}

This triple of partition functions has very similar properties to
the ones considered in the previous section, the $Z^R_{\mathit{NS},g}$ is
integer moded, and $Z^{\mathit{NS},g}_R$, $Z^{\mathit{NS},g}_{\mathit{NS},g}$ will have moding
$q^n, q^{n+1/2}$ in the non anomalous case, $q^{n\pm 1/4}$ in the
anomalous case.
The other six partition functions form a separate, cyclic orbit,
with the NS partition function with g insertion $Z^{\mathit{NS}}_{\mathit{NS},g}$ sent
by T to $Z^{\mathit{NS}}_{R,g}$, sent by S to $Z^{R,g}_{\mathit{NS}}$, sent by T to
$Z^{R,g}_{\mathit{NS},g}$, sent by S to $Z^{\mathit{NS},g}_{R,g}$, sent by T to
$Z^{\mathit{NS},g}_{\mathit{NS}}$, sent by S back to the beginning of the orbit. We
will not make any use of this orbit in the following section.

Let us collect the information available about the polar terms in the non anomalous case. First
of all, in any R sector there should be no polar terms.  Following the example of the bosonic theory one can exclude
twist fields of dimension $1/2$, $1$ in NS
g-twisted sectors and cap the number of fields
with dimension $3/2$ to one, because the OPE of two NS twist fields
gives untwisted fields in the NS sector, which for level smaller
than $3$ are only the identity and e.m. tensor. Furthermore we know that
there is only one Ramond ground state, which is has to be
g-invariant if the Monster is a symmetry of the theory.
The orbit of size three has polar terms
only from the NS g-twisted fields, hence if there are no twist fields
of dimension $3/2$ the partition function will just be constant,
one. Inspection of the characters of the $2A$ and $2B$ classes shows
that this cannot be the case for the Ramond partition function with
$2A$ or $2B$ insertion. If there is one field of dimension $3/2$,
then the resulting partition function actually agrees well with the
result expected from a $2B$ twist, with the coefficients of
$Z^R_{\mathit{NS},g}$ decomposing nicely into the expected $2B$ characters.
Again, the $2A$ twisted partition function cannot be non-anomalous.
To reproduce the expected $2A$ character of dimension 3 Ramond
fields one would need $47$ dimension $3/2$ NS 2A-twisted fields, and
then the rest of the partition function would still not decompose
properly into $2A$ characters.
The anomalous case also does not work, for the same reason as in the
previous section. The constant coefficient in $Z^R_{\mathit{NS},g}$ turns out
to be again $8 (b_{1/4} + 4 b_{3/4} + 6 b_{5/4}$ in terms of the
polar coefficients of $Z^{\mathit{NS},g}_R$. Hence it cannot be one, as it
should be for the ESCFT.
We conclude that the $k^*=4$ ESCFT cannot have monster symmetry.

\section{Size of a chiral algebra vs order of minimal monic
differential constraints}

The author of \cite{Gaberdiel:2007ve} is interested in an argument
of \cite{Zhu:1996} relating a certain algebraic property of a chiral
algebra ${\cal A}$, the size $s$, to the existence of a monic
differential operator of degree $s$ with coefficients in polynomials
of the Eisenstein series $E_2,E_4,E_6$ which annihilates the
characters of all the representations of ${\cal A}$. The size is
defined as the smallest degree of a monic polynomial in $L_{-2}$
which sits in a certain subspace ${\cal O}_q$ of the vacuum
representation ${\cal H}_0$ of ${\cal A}$. If a chiral algebra has
size $s$ then $L_{-2}^s$ will always sit in another important
subspace called ${\cal O}_{[2]}$. ${\cal O}_{[2]}$ is spanned by
vectors of the form $\{A_{-h_A-1} |v\rangle|| A\in {\cal A},
|v\rangle \in {\cal H}_0 \}$.
The important conjecture of \cite{Gaberdiel:2007ve} is essentially
the converse of the above result: the minimum degree of a monic
differential operator which annihilates all the characters of the
chiral algebra should be the same as the size of the algebra. The
conjecture rules out ECFTs for high enough $k$, because the
partition function of ECFTs of high $k$ is annihilated by monic
differential operators of degree smaller than $k$, but the size of
the chiral algebra of an ECFT is bigger than $k$

We can build a simple counterexample to this claim: it is the $n$-th
power of the Monster module. The monster module has size $3$:
$L_{-2}^3$ sits in ${\cal O}_{[2]}$ while $L_{-2}^2$ does not.
Correspondingly the partition function $J(q)$ of the monster CFT is
annihilated by a monic differential operator of degree $3$. The size
of the $n$-th power of the Monster CFT is at least $2n+1$:
$(L^{\mathit{tot}}_{-2})^{2n+1}= (\sum L^i_{-2})^{2n+1}$ sits in ${\cal
O}_{[2]}$, as it is the sum of terms each of which contains at least
one $L^i_{-2}$ raised to the power of $3$ or more, which sits in
${\cal O}_{[2]}$; on the other hand $(L^{\mathit{tot}}_{-2})^{2n} \sim
\prod_i (L^i_{-2})^2 + v, v\in {\cal O}_{[2]}$, which is not in
${\cal O}_{[2]}$. (A similar argument for the theory  $e_8^l$ is given in
\cite{Gaberdiel:2007ve})

By inspection for the first few values of $n$ (up to $n=15$), it is
possible to show that $J^n$ satisfies a monic differential equation
of degree $n+2$ and no monic equation of degree $n+1$. Already for
$n=2$ this degree is smaller than the size. Essentially, the
partition function may satisfy ``accidental'' differential
equations, which are not due to singular vectors in the language of
\cite{Gaberdiel:2007ve}. Hence we see no obstacle to have ECFTs with
arbitrarily high $k$.

\section*{Acknowledgements}
We are grateful to M.Gaberdiel for correspondences, to X. Yin for
helpful discussions and especially to E. Witten for  discussions and
detailed comments on an earlier draft of this paper. The author is
supported by NSF grant PHY-0503584

\appendix
\section{Twisted Hecke transforms}
The most natural way to describe the partition function $Z$ of any
selfdual CFT with $c=24k$ is to use Hecke transforms of the $J$
function, which is the partition function of the Monster CFT,
\cite{Witten:2007kt}, as
\begin{equation}
Z = \sum a_i\,H_i\,J(\tau)
\end{equation}
Here the coefficients $a_i$ are the coefficients of the polar terms.
This means that the coefficients of the partition function of any
selfdual CFT with $c=24 k$ decompose naturally in dimensions of
Monster group representations. Clearly the Monster is not the
symmetry group of all selfdual CFTs, and one might believe that the
existence of a consistent set of Monster-twisted partition functions
would be a much stronger requirement, able to discriminate
candidates with Monster symmetry.
In this Appendix we mean to show that a concept of "twisted Hecke
transforms" exists, which automatically builds a set of $Z_{g}$
which are consistent with $Z$ and with each other.

For a physicist, the Hecke transforms of the partition function $F$
of a CFT, $H_n F(\tau)$, are defined as the building blocks for the
partition function of symmetric product orbifolds of that CFT. The
chiral partition functions for a chiral symmetric product of a
selfdual CFT can be computed by the methods of
\cite{Dijkgraaf:1996xw} and collected in a generating function
\begin{equation}
U(p,q)=\sum_N p^N Z^{\mathit{Sym} N}(\tau) = \frac{1}{\prod_{n>0} (1-p^n
q^m)^{c(nm)}}
\end{equation}
Here $F(\tau)=Z^1(\tau) = \sum c(n) q^n$.
The logarithm of the generating function is the generating function
for the Hecke transforms of $F$.
\begin{equation}
\log U(p,q) = \sum_N \frac{1}{N} H_N F(\tau)
\end{equation}

Notice that the symmetric orbifold of $n$ copies of the Monster CFT
still possess a diagonal Monster symmetry. The partition function of
the symmetric product can be twisted by this symmetry, and the
resulting set of partition functions will clearly be self-consistent
(As they are the partition functions of an actual theory). We can
collect then into a generating function computed in the spirit of
\cite{Dijkgraaf:1996xw} (See also appendix D of \cite{David:2006yn}
for an interesting example).
The result is
\begin{equation}
U_g(p,q)=\sum_N p^N Z^{\mathit{Sym} N}_g(\tau) = \frac{1}{\prod_{n>0}
\det_{{\cal H}(nm)} (1-p^n q^m g)}
\end{equation}
Here  $Z^1_g = \sum q^n\,\mathrm{Tr}_{{\cal H}(n)}\,g $.
The twisted Hecke transforms can be extracted from the logarithm, as
\begin{equation}
(H_N F)_g = \sum_{d|N} \sum_{b=0}^{d-1} F_{g^{N/d}}\left(\frac{N
\tau + b d}{d^2}\right) \end{equation}
and
\begin{equation}
(H_N F)^{g^a}_{g^c} = \sum_{d|N} \sum_{b=0}^{d-1} F^{g^{a d}}_{g^{c N/d-a b}}\left(\frac{N
\tau + b d}{d^2}\right) \end{equation}
It is easy to check that $(H_N F)^{g^a}_{g^c}$ have the correct modular transformation rules.

It is quite clear by induction over $k$ that the twisted Hecke
operators of the $k=1$ twisted partition functions $(H_k t)_g$ provide good building blocks for the twisted
partition functions of an hypothetical ECFT with Monster symmetry
$Z_{e,g}$.
\begin{equation}
Z_{g} = \sum a_i (H_i t)_{e,g}
\end{equation}
This has the appropriate $q$ expansion, with coefficients which
decompose into characters $\chi_R(g)$ in a way that is compatible
with the decomposition of the coefficients of $Z = \sum a_i H_i
J(\tau)$ in dimensions $d_R$.
Moreover the modular transforms
\begin{equation}
Z_{g^a,g^b} = \sum a_i (H_i t)_{g^a,g^b}
\end{equation}
also have by construction coefficients which decompose into
dimensions of irreps of the centralizer of $g$, with coefficients
depending on the numbers $a_i$, which are pretty much as natural as
possible.

\bibliography{strings}{}
\bibliographystyle{plain}

\end{document}